\documentstyle[11pt,aaspp]{article}
\def\gs{\mathrel{\raise0.35ex\hbox{$\scriptstyle >$}\kern-0.6em %
\lower0.40ex\hbox{{$\scriptstyle \sim$}}}}
\def\ls{\mathrel{\raise0.35ex\hbox{$\scriptstyle <$}\kern-0.6em % Less
\lower0.40ex\hbox{{$\scriptstyle \sim$}}}}

\def\arcsper{\ifmmode \rlap.{''}\else $\rlap{.}''$\fi}
\def\arcmper{\ifmmode \rlap.{'}\else $\rlap{.}'$\fi}

\def\VI{\hbox{$(V\! -\! I)$}}
\def\RI{\hbox{$(R\! -\! I)$}}
\def\VR{\hbox{$(V\! -\! R)$}}
\received{---}
\revised{---}
\accepted{---}

\begin{document}
\small

\title{DEEP OPTICAL GALAXY COUNTS WITH THE\\
 KECK TELESCOPE\footnote{Based on observations
obtained at the W.M.\ Keck Observatory,
which is operated jointly by the California Institute of Technology
and the University of California.}}
\author{
Ian Smail\altaffilmark{1},
David W.\ Hogg\altaffilmark{2},
Lin Yan\altaffilmark{3}
\& Judith G.\ Cohen\altaffilmark{3}.
}
\affil{\tiny 1) The Observatories of the Carnegie Institution of Washington,
813 Santa Barbara St., Pasadena, CA 91101-1292}
\affil{\tiny 2) Theoretical Astrophysics, Caltech 130-33, Pasadena CA 91125}
\affil{\tiny 3) Palomar Observatory, Caltech 105-24, Pasadena CA 91125}

\begin{abstract}
We present faint galaxy counts from deep
$VRI$ images obtained with the Keck Telescope.
These images reach $R\sim27$ in median seeing FWHM $\sim 0.5$--0.6 arcsec
and we detect a integrated galaxy number density of $7\times 10^{5}$
degree$^{-2}$, equivalent to $3\times 10^{10}$ galaxies in the
observable Universe. In addition we present median galaxy colors as a
function of magnitude; bluing trends are visible in all colors to
$R\sim 24.5$.  Fainter than $R\sim24.5$, however, the typical
\VR\ color becomes redder again, \VI\ remains constant, and
\RI\ becomes yet bluer.  These trends are consistent with the $VRI$
count slopes, implying a decrease in the $V$ slope at the faintest
levels, which our data supports.  Taking advantage of our good seeing
we also present median half-light radii for faint galaxies,
these show a steady decline at fainter magnitudes, leading to an
intrinsic half-light radius of $\sim 0.2$ arcsec for a typical $R\sim25.5$--26
galaxy.  Irrespective of the redshift distribution, the extremely high
galaxy surface densities and their small intrinsic sizes are consistent with
a scenario in which the majority of the very faint field population
are dwarf galaxies or sub-galactic units.
\end{abstract}


\sluginfo

%\vfil\eject

\section{Introduction}

The issue of the total number of galaxies in the visible Universe is
more than an academic curiosity; number counts as a function of
apparent magnitude hold important clues to the formation and evolution
of galaxies (see \cite{koo92} for a review).  The study of very faint
field galaxies using CCDs was pioneered by Tyson (1988) who found an
integrated number density of $\sim 2\times 10^{5}$ degree$^{-2}$, or
$\sim 10^{10}$ galaxies over the whole sky at $B\sim26$.  Two more
recent studies which are notable for their depth and resolution are
those of the Hawaii group (\cite{lilly91}) and the Durham group
(\cite{metcalfe91}, \cite{metcalfe95}).  The latter now reach galaxy
surface densities in excess of $4\times 10^5$ degree$^{-2}$ for a
small field at a limiting magnitude of $B=27.5$. Equivalent surface
densities are also being achieved in the near-infrared, $K$
(\cite{gardner93}, \cite{cowie95}, \cite{sgd95}).

Potentially the most interesting result from these studies, first
observed by Lilly et al.\ (1991) and confirmed by Metcalfe et
al.\ (1995) with deeper data, is the change in the slope, $\alpha$
(where $\log dN/dm\sim\alpha\,m$), of the deep $B$ counts.  The
differential galaxy number counts in the $B$-band show a decrease in
slope from $\alpha_B\sim 0.5$ to $\alpha_B\sim 0.3$ at $B\sim25$
(equivalent to $R\sim 24$--24.5).  In addition, Roche et al.\ (1993)
claim that the amplitude of the two-point angular correlation function
reaches a minimum around $B\sim25$.  Taken together these two
observations have been interpreted by Metcalfe et al.\ (1995) as a
signature of an increasing dominance of low luminosity galaxies at
$z\gs 1$.  These features, if confirmed and observed in other
passbands, will constrain galaxy evolution models, although redshift
information will be required before any strong conclusions can be
drawn.

This letter presents the first deep optical counts with the 10-m Keck
Telescope.  These $VRI$ observations are both deeper, at a given
completeness limit, than previous published observations
and more significantly were obtained in good seeing, FWHM $\sim
0.5$--$0.6$ arcsec, providing a wealth of information on the very faint
field population.

%\vfil\eject

\section{Observations and Analysis}

\subsection{Observations and Reduction}

The observations presented here were obtained by S.R.~Kulkarni and
J.G.~Cohen to study the optical emission from two high galactic
latitude pulsars.  Thus for our purposes the fields provide randomly
selected samples of the faint field population.  These data were
obtained on the night of 1994 August 8 using the Low Resolution Imaging
Spectrograph (LRIS, \cite{cohen95}) on the 10-m Keck Telescope, Mauna
Kea. The imaging mode of LRIS provides a $5.7\times 7.3$ arcmin field
onto a thinned Tek detector with 0.21 arcsec/pixel sampling and
8.0$e^-$ readnoise.  Conditions during this run were good and the night
was photometric; a log of the observations is given in Table~1.  The
raw frames, with individual exposure times between 0.5--1.5ks, were
debiased and then flatfielded in a standard manner using dome
flatfields.  For each passband the  dithered frames, with pointing
offsets of 10--20 arcsec, were registered and combined using standard
IRAF routines and a clipped-average algorithm to produce the final
images.  We reproduce a randomly chosen $1\times1$ arcmin region from
the $R$ frame of 1640+22 in Figure~1.

Photometric calibrations were derived from observations of Landolt
(1992) standard stars at air masses very close to those of our science
observations. We estimate the average errors in our absolute calibrations
to be $\delta \ls 0.02$. Color differences between the standard
stars and typical faint galaxies give systematic errors of a similar
size.  The estimated reddening is $E(B\!  -\! V) = 0.07$
(\cite{stark92}) for both fields and so we have applied zero point
corrections of $\delta V = 0.16$, $\delta R = 0.12$ and $\delta I =
0.07$.  We conservatively estimate our absolute magnitude calibrations
to be good to $\ls 0.1$ mag.

\subsection{Analysis and Photometry}

To analyse these images we elected to use the SExtractor image analysis
package (\cite{bertin94}).  This package is a  fast, image analysis
program which can detect (using a standard  isophotal limit and  area
cut algorithm),  robustly deblend and analyse sparse and moderately
crowded galaxy fields.  The program provides positions, shapes, image
profiles, half-light radii, isophotal and aperture magnitudes for all
deblended objects detected on a frame.  To detect objects we first
smooth the frame with the point spread function,   then
thresholded it at a level of $2.5 \sigma$ of the sky noise in the raw
frame and identify objects. This threshold provides a typical point
source detection limit of $R \sim 27.2$ and appeared optimal in our
incompleteness and noise object analyses.

To obtain total magnitudes for our galaxies we have followed the
approach used by Djorgovski et al.\ (1995).
For the fainter objects we use a fixed aperture (1.5 arcsec diameter or
$\sim 2$--$3\times$FWHM) to measure magnitudes, these are then aperture
corrected to a 5.6 arcsec diameter under the assumption that the
objects have roughly stellar profiles, a correction of $\delta \sim -0.14$
at $R=26$.  This assumption was tested by
stacking many faint objects to make ``average'' faint galaxies with
high enough signal-to-noise for large aperture photometry; we find that
the average galaxies have aperture corrections close to those of the
stars (as expected from their typical sizes, see below).  For the
brighter objects in the field, those with isophotal diameters larger
than our adopted fixed aperture, we measure isophotal magnitudes above
a surface brightness threshold of 0.9$\sigma$ of the sky noise
(Table~1) and similarly correct to 5.6 arcsec diameter.   The majority
of the objects with magnitudes brighter than $R\sim20.5$ are saturated
in our frames; we do not present counts brighter than this limit.  To
measure colors for our objects we smooth all the frames to the
effective seeing of the worst frame (the $V$ data), before measuring
photometry in 1.5 arcsec diameter apertures for all the objects
selected from the $R$ catalogs.

Removal of stars from our object catalog is necessary at brighter
magnitudes.  With the good seeing experienced during these observations
it is relatively straightforward to use image concentration to identify
the stellar locus to a depth of $R=24.5$.  Comparison of the colors of
these candidate stars to those in Landolt (1992) confirms their
identification and supports our estimates of the errors on our absolute
color calibrations of $\ls 0.1$ mag.  At $R=24.5$ the stellar fraction
is $\ls 6\%$ and falling, so we do not apply any correction for stellar
contamination fainter than $R=25$.

%\vfil\eject

\subsection{Completeness Modelling}

Non-detection,  failure to detect an object that is really there;
false detection,  detection of non-real, noise objects; and magnitude
errors,  incorrect measurement of an object's flux, all plague faint
galaxy counts.  These problems become progressively worse at  fainter
magnitudes, with their relative importance depending upon the slope of
the counts (because magnitude errors ``scatter'' plentiful faint
objects into brighter magnitude bins; Eddington bias) and the details
of the data reduction procedures.

We assess non-detection and magnitude errors simultaneously with a
Monte Carlo simulation that involves adding artificial galaxies to the
data and re-applying the detection algorithm.  In order to simulate
both the mean properties and morphological diversity of faint galaxies
we create artificial objects by extracting galaxy images at magnitude
levels for which  errors are negligible and dimming them by 2.5
magnitudes, at a constant angular size.  It should be noted that if the
scale sizes of galaxies decrease at fainter magnitudes (c.f.\ \S 3)
then this procedure will tend to overestimate our incompleteness.
Extensive simulations ($>10^4$ objects per frame, dropped in one at a
time) were used to create a large matrix $P_{ij}$ each element of which
is the probability that a galaxy with magnitude $m_j$ is detected with
a measured magnitude $m_i$.  This matrix contains all of the
completeness and magnitude error information.  On the assumption that
the counts do not turn over quickly (i.e.\ $\alpha$ does not change
rapidly with magnitude near the limit of the data), $P_{ij}$ can be
projected into a total detection rate $r_i$, the ratio of number
observed to true number at magnitude $m_i$.  The true number of
galaxies in bin $i$ can then be estimated by dividing the observed
number $N_i$ by the detection rate $r_i$.

To correct for false detections we have created noise frames identical
to our observations, geometrically remapped, combined and analysed
them in the same manner as the observations.  We subtract the number
counts of noise objects found in these simulated frames from our
observed counts.  This correction is a few percent or less, even in our
faintest magnitude bins.

The faintest magnitude at which the counts can be reasonably
completeness-corrected is subjective; we truncate our counts at or
before the $50\%$ completeness level, beyond which the Poisson errors,
or any other error estimates, lose their meaning.

\section{Results and Discussion}

We plot in Figure~2 the raw and corrected differential galaxy number
counts from our two fields.  We measure slopes from the combined field
counts of:  $\alpha_V = 0.404\pm 0.015$ for $V=22$--$24.25$, flattening
to $\alpha_V = 0.28\pm0.05$ at fainter magnitudes, $\alpha_R =0.321\pm
0.001$ for $R=21$--$27$ and $\alpha_I =0.271\pm 0.009$ for
$I=19.5$--$25.5$. We see no obvious variation in the $R$ and $I$
slopes. The evidence for the break in $V$ count slope at $V\sim 24.5$
is certainly not pronounced, but it is stronger in the deeper,
wider-area 1640+22 data, and as we show below it is also seen in the
median galaxy colors.  The integrated, corrected number counts are
$2.9\times 10^5$ per sq.\ degree to $V=26.5$, $7.3\times 10^5$ to
$R=27$, and $4.1\times 10^5$ to $I=25.5$.  We also plot on Figure~2
counts by other workers.  The discrepencies in the count slopes at
bright magnitudes may arise from other groups  choosing
fields devoid of bright galaxies,  this biases them towards underdense
regions producing a systematic undercount of galaxies at bright and
intermediate magnitudes and thus a steeper slope to the counts.  At
faint limits we find reasonable agreement between the slopes of the
various datasets, with residual differences in normalisation consistent with
the different photometric corrections applied by different workers and
our quoted photometric errors.

The color distributions for our faint galaxy sample are shown in
Figure~3. All show a bluing trend with fainter magnitude
until $R\sim 24$--24.5,  where a typical
galaxy has a flat spectral energy distribution in $\nu\,f_{\nu}$
(corresponding to $(V-R) \sim 0.5$ and $(R-I) \sim 0.5$).
Fainter than $R\sim 24$--24.5 the typical \VR\ colors turn-around and
become redder, while the \VI\ colors flatten out and the \RI\ continue
to become bluer (c.f.\ \cite{steidel93}).   If we are observing the
same population in all three bands, then these trends in median color
should be reflected in different count slopes in $VRI$.  At $R\ls24$
the bluing trends indicate $\alpha_V > \alpha_R > \alpha_I$ as observed.
After the turn-around, the colors imply a decrease in $\alpha_V$
from steeper than $I$ at the bright end, to matching the $I$ slope at
the faint end.  This change in $V$-band slope is indeed seen in the
counts, although its significance is not high.

The good seeing of our observations allow us to study the angular sizes
of the faint galaxy population.  We plot in Figure~4 the observed
half-light radii, $r_{hl}$, measured from the $R$-band images.  A
gradual, near linear, decline is seen in the median size of faint
galaxies to $R\sim25.5$--26, at which point the median size is only
just distinguishable  from the stellar locus.  Representing the faint
field population as exponential disks leads to an estimate of the
intrinsic half-light radius of a typical $R=25.5$--26 galaxy of $r_{hl}
\sim 0.2$ arcsec.  This angular size corresponds to less than
$2.9\,h_{50}^{-1}~{\rm kpc}$ at {\it any} redshift in {\it any} world
model, so it is clear that the typical objects we are detecting are either
intrinsically smaller or produce a larger fraction of their emission in
their nuclear regions than local bright galaxies.  Splitting the sample in
two in each magnitude bin on the basis of \VI\ color shows no
statistical difference between the sizes of the blue and red galaxies.

In conclusion, we have presented deep galaxy counts in $VRI$
passbands.  We find a decrease in slope in our bluest passband, $V$, at
the faint end similar to that seen in the $B$-band by Metcalfe et al.\
(1995) at an equivalent apparent magnitude.  Beyond $R\sim 24$ it
appears that galaxy counts approach $\log dN/dm\sim 0.3\,m$ in all
bands, shallow enough to remove the threat of a divergence in the
integrated night sky brightness.  If we are not seeing to very high
redshift, either because the Universe is Einstein-de Sitter and we are
running out of volume, or because there are few luminous, high redshift
field galaxies, then the common faint-end slope in the galaxy counts is
a measure of the shape of the galaxy (or possibly pre-galactic
fragment, in view of the small angular size of the faintest objects)
luminosity function at the faint end.  In these scenarios, the slope of
$0.3$ implies a faint end luminosity function $\Phi(L)\sim L^{-1.75}$
at large look-back times, in contrast to $L^{-1}$ observed locally
(e.g.\ \cite{loveday92}).  The faint counts may thus confirm, at higher
redshift, the steepening of the faint-end slope of the luminosity
function suggested by analyses of spectroscopic samples at brighter
magnitudes (\cite{eales}, \cite{rse}).  Interestingly, the roll-over at
$V\sim 24.5$ is within reach of the new generation of large telescopes
and therefore we may hope for direct spectroscopic observations of
galaxies in this intriguing magnitude range.

\section*{Acknowledgements}

Firstly, we thank Shri Kulkarni for his great generosity in allowing us
to use these data.  We  acknowledge useful discussions and
encouragement from Rebecca Bernstein, Roger Blandford and Nigel
Metcalfe.  We also thank the referee, Richard Kron, for many
helpful comments.  Support via a NATO Advanced Fellowship and a
Carnegie Fellowship (IRS) and an NSF Graduate Fellowship (DWH) is
gratefully acknowledged.  Finally, it is a pleasure to thank the
W.M.\ Keck Foundation and its President, Howard B.\ Keck, for the
generous grant that made the W.M.\ Keck Observatory possible.

\smallskip

\section*{Tables}

\noindent{\bf Table 1} The table gives the log of the observations
including the field identification and coordinates.  For each filter we
also quote the total exposure time, $T_{exp}$; the total number of sky
photons detected per pixel, $N_{\gamma}$; the 1$\sigma$ surface
brightness limit, $\mu(1\sigma)$; the 50\% completeness limit from our
simulations, $m_{lim}$; the number of objects detected above this limit
on the frame, $N_{lim}$, the FWHM of the seeing in arcsec and the total
field area in sq.\ arcmin.

\section*{Figures}

\noindent{\bf Figure 1}
A randomly selected $1 \times 1$ arcmin field from the $R$ image of
1640+22.  Objects with magnitudes in the range $R=26.0$--26.5 are
marked.  Notice the minimal crowding in the frame.

\noindent{\bf Figure 2}
Plots of differential galaxy counts as a function of magnitude in the
$VRI$ passbands in the two fields.  The raw and
corrected counts are represented with skeleton and solid symbols
respectively, circles for field 1640+22, squares for 2229+26.  The
error bars are Poisson plus (on the solid points) an estimated
uncertainty in the completeness correction.  Least-square linear
fits are shown with solid lines (there are two fits to the $V$
counts split at $V=24.25$).  Where applicable, the work of other
authors is shown with open points, these are truncated at the
same completeness limit as our data.  The following corrections
have been applied to transform the different passbands onto
our system:  Lilly et al.\ (1991) $I = I_{\rm AB} - 0.48$ and
Steidel \& Hamilton (1993)  $R = {\cal R}_{\rm AB} + 0.16$.

\noindent{\bf Figure 3}
The typical colors of the faint galaxy population as a function of $R$
magnitude in the two fields.  The points are medians of samples of
401 galaxies in each bin.  The horizontal error-bars show the extent
of the magnitude bin, while the vertical error-bars are 95\%
confidence limits calculated using boot-strap resampling of the data,
non-detections are included.  $1\sigma$ limits within our photometry
aperture lie outside the region of the color planes plotted.
Note the turn-around in the \VR\ colors at $R\sim24.0$--24.5 and the
flattening of the \VI\ colors at the same magnitude.

\noindent{\bf Figure 4}
The half-light radii as a function of apparent $R$ magnitude for all
the objects in our two fields.  The points are medians of samples of
401 galaxies with error-bars defined in a similar manner to Figure~3.
The seeing difference between the fields has been removed by shifting
the 2229+26 points.  The dotted line marks the locus of stars, these
are excluded from the medians.   Each of the dashed lines is a linear
fit to the variation of apparent half-light radius with source
magnitude for an exponential disk with a fixed, intrinsic half-light
radius, labelled in arcsec. The solid line shows the detection limit as
a function of half-light radius for an exponential disk.  For the most
extended sources ($r_{hl} \gs 0.6$ arcsec) a bias is visible, such that
the source sizes are progressively underestimated at fainter
magnitudes.  This bias arises from a combination of incompleteness
beyond $R\sim25$ and the use of inappropriate aperture corrections,
the  corrections  applied are calculated for  ``typical'', hence
compact, faint galaxies.  Nevertheless, it is apparent that the median
source size in the field shrinks more rapidly than expected for a
population with a fixed intrinsic angular size.  By $R\sim25.5$--26 the
half-light radii is asymptotically approaching the stellar locus and
indicates a very small intrinsic size for these faint galaxies, $r_{hl}
\sim 0.2$ arcsec.
\smallskip

\cleardoublepage
\moveright -0.5truein\vbox{
%\begin{center}
\begin{tabular}{lcccccccccc}
\multispan9{\bf \hfil \hfil Table 1 \hfil }\\
\noalign{\smallskip}
\hline
\noalign{\smallskip}
Field & $\alpha$ & $\delta$ & Filter & $T_{exp}$ & $N_{\gamma}$ & $\mu
(1\sigma)$ &  $m_{lim}$ & $N_{lim}$ & FWHM & Area \\
\hfil & J2000 & J2000 & \hfil & ks & $10^3 e^-$ & mag/\sq $''$ & 50\% & 50\% &
$''$ & \sq $'$ \\
\noalign{\smallskip}
\hline
\noalign{\smallskip}
1640+22 & $16^h 40^m 18\arcsper 90$ & $+22^\circ 24' 19\arcsper 0$ & V & 1.5 &
16.5 & 28.31 & 27.1 & 4562 & 0.78 & 39.8\\
\hfil   & $l=41.1$ & $b=38.3$   & R & 2.4 & 44.3 & 28.49 & 26.9 & 5611 & 0.55 &
40.2\\
\hfil   & \hfil & \hfil   & I & 2.0 & 164.1 & 27.06 & 26.0 & 4618 & 0.53 &
40.3\\
\noalign{\smallskip}
2229+26 & $22^h 29^m 50\arcsper 89$ & $+26^\circ 43' 52\arcsper 8$ & V & 0.9 &
9.0 & 28.33 & 26.6 & 1306 & 0.87 & 15.0\\
\hfil   & $l=87.7$ & $b=-26.3$   & R & 2.1 & 41.3 & 28.05 & 26.9 & 5254 & 0.58
& 40.8\\
\hfil   & \hfil & \hfil   & I & 1.0 & 94.6 & 26.34 & 25.6 & 3284 & 0.58 &
36.0\\
\hline
\noalign{\smallskip}
\noalign{\smallskip}
\hline
\noalign{\smallskip}
\end{tabular}}
%\end{center}

\vspace*{.5cm}


\begin{thebibliography}{}
\itemsep=0in

\bibitem[Bertin 1994]{bertin94} \reference
Bertin E., 1994, SExtractor manual, IAP, Paris.

\bibitem[Cowie et al.\ 1995]{cowie95} \reference
Cowie L.L., Gardner J.P., Hu E.M., Songaila A., Hodapp K.W. \&
Wainscoat R.J., 1995, ApJ 434, 114

\bibitem[Djorgovski et al.\ 1995]{sgd95} \reference
Djorgovski S., Soifer B.T., Pahre M.A., Larkin J., Smith J.D.,
Neugebauer G., Smail I., Matthews K., Hogg D.W., Blandford
R.D., Cohen J., Harrison W. \& Nelson J., 1995, ApJ 438, L13

\bibitem[Eales 1993]{eales} \reference Eales, S., 1993,
ApJ, 404, 51

\bibitem[Ellis et al.\ 1995]{rse} \reference Ellis R.S., Colless M.M.,
Broadhurst T.J.,  Heyl J.S.\ \& Glazebrook K., 1995, MNRAS, submitted.

\bibitem[Gardner et al.\ 1994]{gardner93} \reference Gardner J.,
Cowie L.L. \& Wainscoat R.J., 1993, ApJ 415, L9

\bibitem[Koo \& Kron 1992]{koo92} \reference Koo D.C. \& Kron R.G.,
1992, ARA\&A 30, 613

\bibitem[Landolt 1992]{landolt92} \reference Landolt A.U., 1992,
AJ 104, 340

\bibitem[Lilly et al.\ 1991]{lilly91} \reference Lilly S.J.,
Cowie L.L. \& Gardner J.P., 1991, ApJ 369, 79

\bibitem[Loveday et al.\ 1992]{loveday92} \reference Loveday J.,
Peterson B.A., Efstatiou G. \& Maddox S.J., 1992, ApJ 390, 338

\bibitem[Metcalfe et al.\ 1991]{metcalfe91} \reference Metcalfe N.,
Shanks T., Fong R. \&  1991, MNRAS 249, 498

\bibitem[Metcalfe et al.\ 1995]{metcalfe95} \reference Metcalfe N.,
Shanks T., Fong R. \& Roche N., 1995, MNRAS 273, 257

\bibitem[Oke et al.\ 1995]{cohen95} \reference Oke J.B., Cohen J.G.,
Carr M., Cromer J., Dingizian A., Harris F.H., Labreque S.,
Lucinio R., Schaal W., Epps H. \& Miller J., 1995, PASP, in press

\bibitem[Roche et al.\ 1993]{roche93} \reference Roche N.,
Shanks T., Metcalfe N. \& Fong R., 1993, MNRAS 263, 368

\bibitem[Stark et al.\ 1992]{stark92} \reference Stark A.A.,
Gammie C.F., Wilson R.W., Bally J., Linke R.A., Heiles C.\ \&
Hurwitz M., 1992, ApJS, 79, 77

\bibitem[Steidel \& Hamilton 1993]{steidel93} \reference
Steidel C.C. \& Hamilton D., 1993, AJ 105, 2017

\bibitem[Tyson 1988]{tyson88} \reference Tyson J.A., 1988, AJ 96, 1

\end{thebibliography}
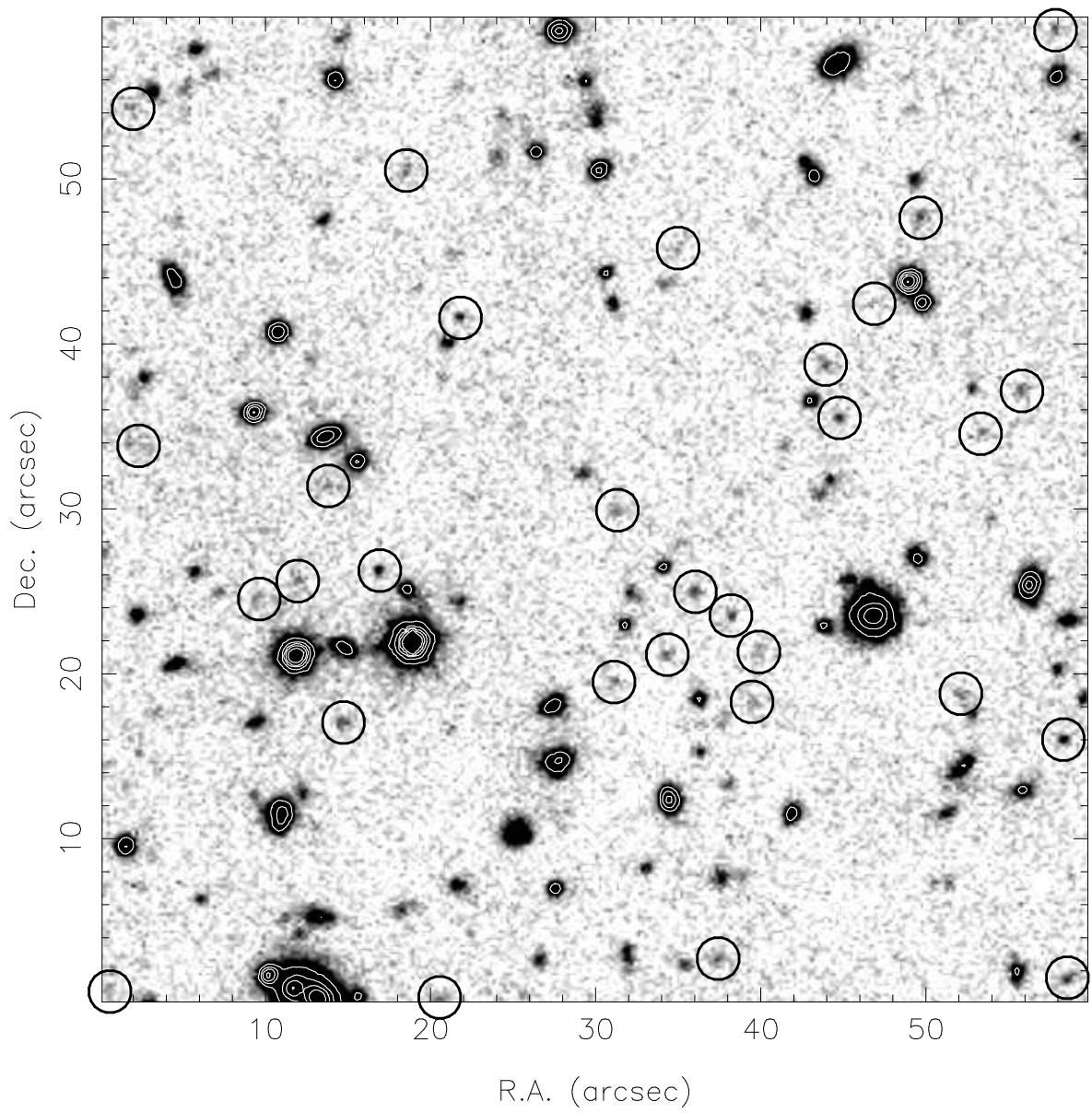
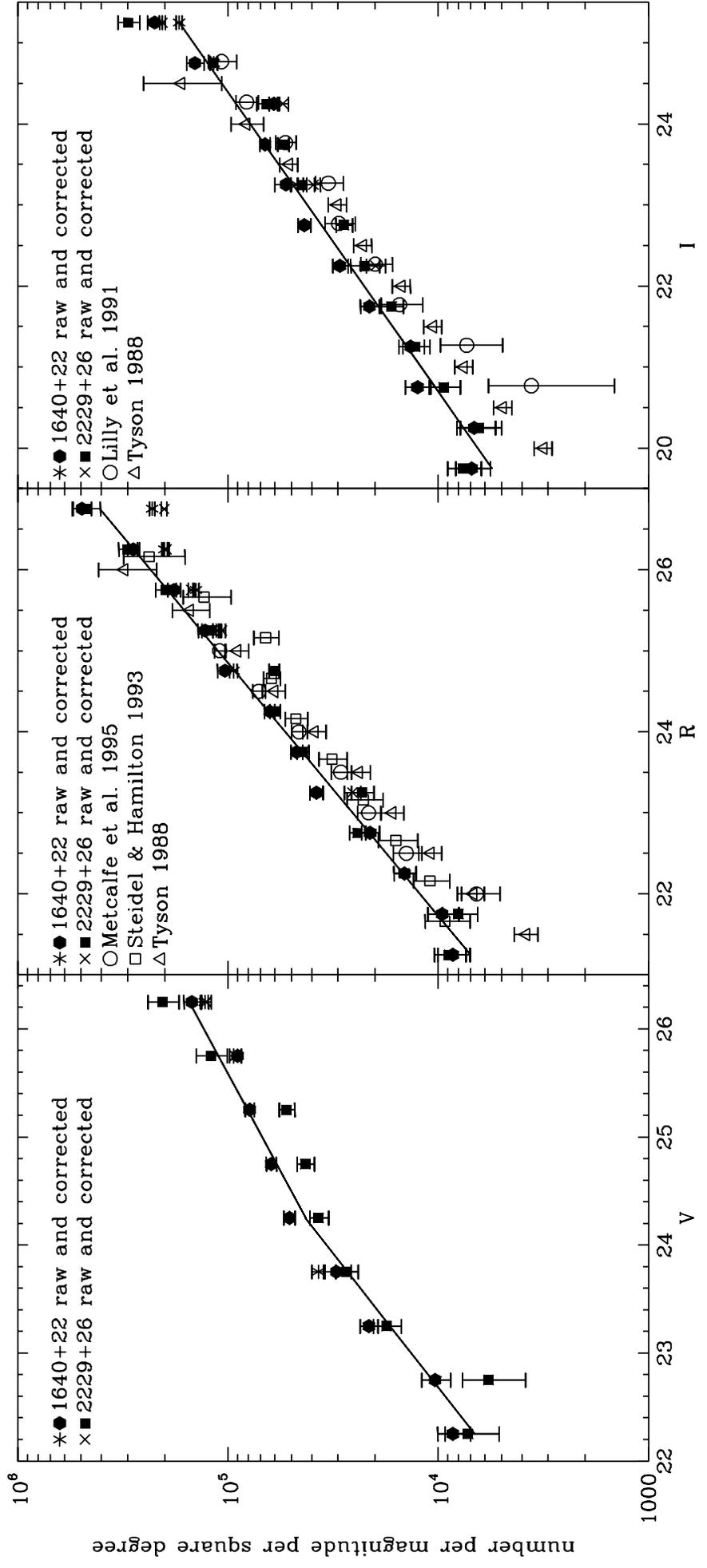
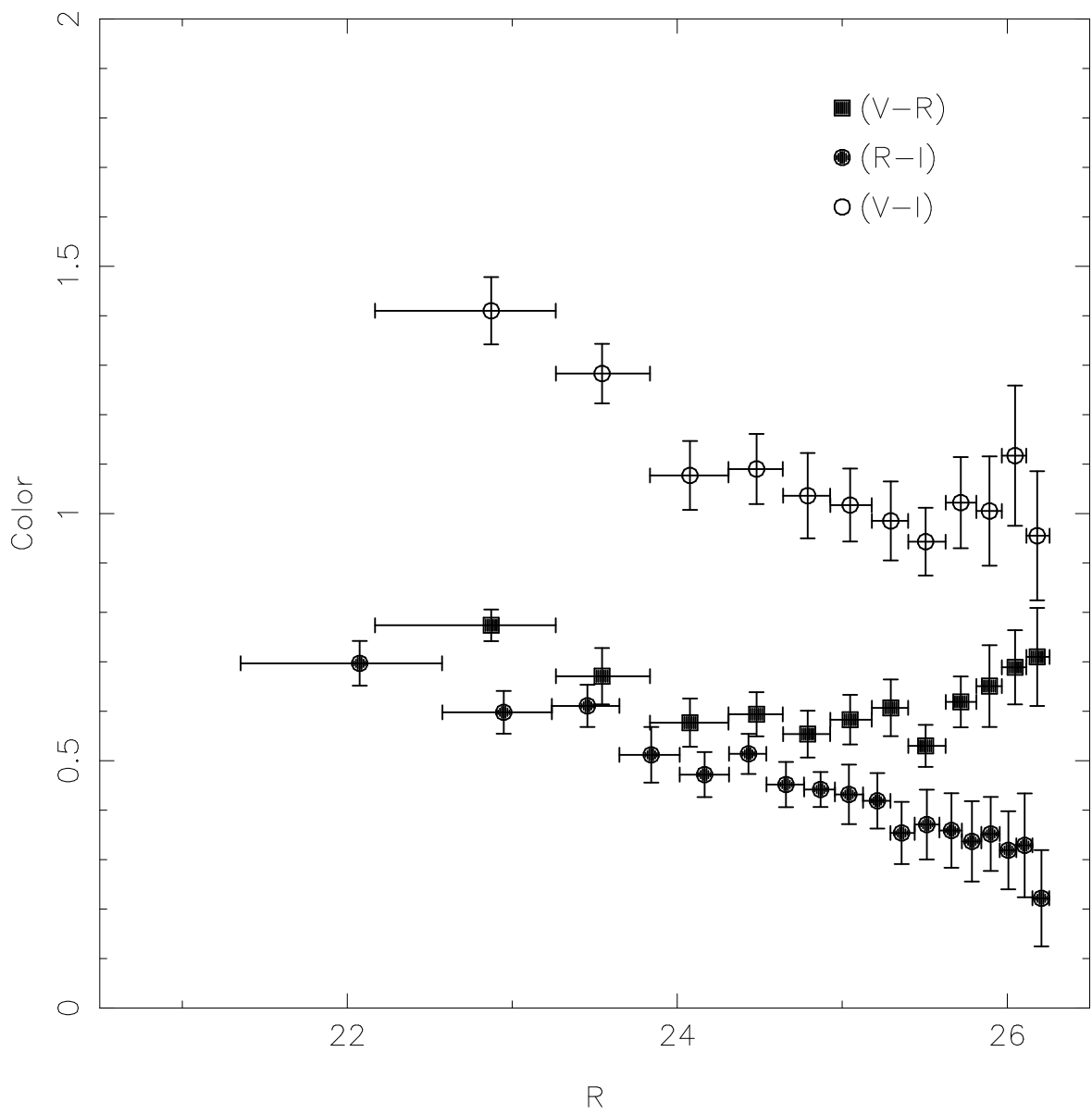
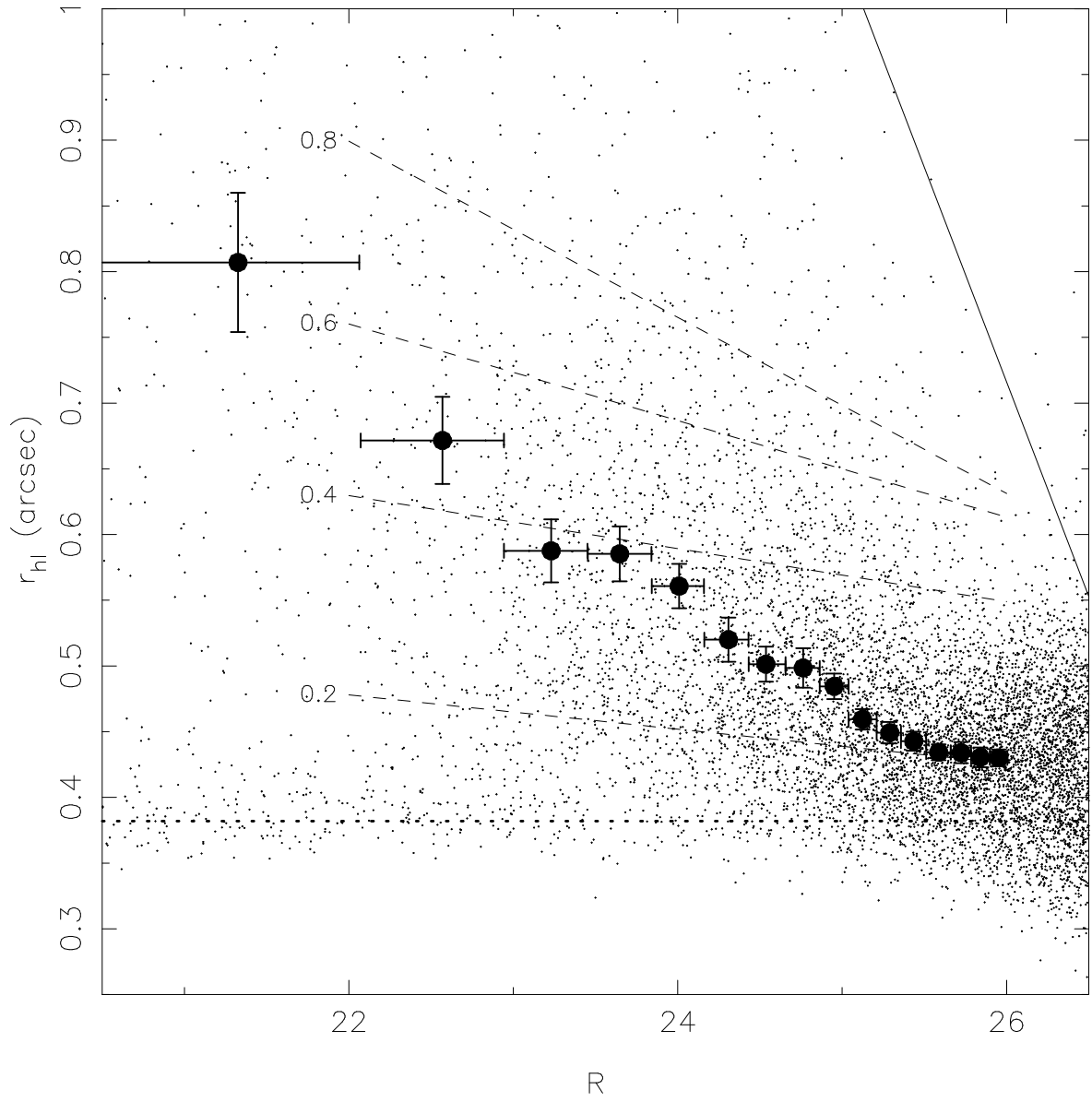
\end{document}